\documentclass[conference,letterpaper]{IEEEtran}
\IEEEoverridecommandlockouts
\usepackage{cite}
\usepackage{amsmath,amssymb,amsfonts}

\usepackage{textcomp}
\usepackage{xcolor}
\def\BibTeX{{\rm B\kern-.05em{\sc i\kern-.025em b}\kern-.08em
    T\kern-.1667em\lower.7ex\hbox{E}\kern-.125emX}}

\usepackage{ textcomp }
\usepackage{multirow}
\usepackage{longtable}
\usepackage{amssymb}
\usepackage{graphicx}
\usepackage{xcolor}
\usepackage{algorithm,algorithmic}
\usepackage[algo2e]{algorithm2e}
\begin{document}

\title{Deep Learning Based Spatial User Mapping on Extra Large MIMO Arrays}
\author{
\IEEEauthorblockN{Abolfazl Amiri, Carles Navarro Manch\'on, Elisabeth de Carvalho  }\\
\IEEEauthorblockA{ Department of Electronic Systems, Aalborg University, Denmark\\
Email: \{aba,cnm,edc\}@es.aau.dk}}

\maketitle

\begin{abstract}
In an extra-large scale MIMO (XL-MIMO) system, the antenna arrays have a large physical size that goes beyond the dimensions in traditional MIMO systems. Because of this large dimensionality, the optimization of an XL-MIMO system leads to solutions with prohibitive complexity when relying on conventional optimization tools. In this paper we propose a design based on machine learning for the downlink of a multi-user setting with linear pre-processing, where the goal is to select a limited \textit{mapping area} per user, i.e. a small portion of the array that contains the beamforming energy to the user. We refer to this selection as \textit{spatial user mapping} (SUM). Our solution relies on learning using deep convolutional neural networks with a distributed architecture that is built to manage the large system dimension. This architecture contains one network per user where all the networks work in parallel and exploit specific non-stationary properties of the channels along the array. Our results show that, once the parallel networks are trained, they provide the optimal SUM solution in more than $80\%$ of the instances, resulting in a negligible sum-rate loss compared to a system using the optimal SUM solution
while providing an insightful approach to rethink these kinds of problems that have no closed-form solution. 

%Adding more antenna elements on the base station of a massive multiple-input multiple-output (MIMO)system enriches the propagation environment leading to gain higher data-rates and spatial resolution. This new system is called \textit{extra large scale MIMO array (XL-MIMO)} and supports services for beyond 5G applications. However, it introduces new challenges such as higher computational complexity and processing delay. In this paper, we propose a spatial user mapping technique using deep convolutional networks to keep the complexity limited. These networks are employed in a novel distributed way to be able to process in parallel reducing implementation costs and the delay. Numerical results confirm the near-optimal performance of the method. We aim to provide insight into the potential of fusion of deep learning and XL-MIMO systems.
\end{abstract}

%\begin{IEEEkeywords}
%Massive MIMO, Deep learning, beyond 5G, Distributed processing, Large intelligent surface
%\end{IEEEkeywords}
\section{Introduction}
Massive multiple-input multiple-output (MIMO) antenna systems are one the most promising solutions for fulfilling the demands for upcoming fifth generation (5G) wireless mobile communication systems \cite{petar5G}. While first pilots of 5G are being launched, there is a lot of attention among researchers for \textit{beyond-5G} topics \cite{beyond_mimo}. One of the most popular research avenues is that of extending the amount of antenna elements present in base stations (BS) beyond typical massive MIMO configurations, which has been referred to in the literature as \textit{extra large scale MIMO array (XL-MIMO)} \cite{xlmimo_mag} or \textit{large intelligent surfaces} \cite{lis_pos}.

%Adding more antennas in the base station (BS) which leads to new technologies known as \textit{extra large scale MIMO array (XL-MIMO)} \cite{xlmimo_mag} or \textit{large intelligent surfaces} \cite{lis_pos} is one viable proposal and an open research topic.

One viable way to realize XL-MIMO systems is to use the infrastructure of large venues like airports, stadiums, etc. in order to place antenna elements. They can be deployed all in one location or distributed across a large area. Having enormous number of antennas provides extreme spatial resolution that can be used to boost throughput and spectral efficiency, support a larger number of users and extend the wireless service coverage area.

One of the main challenges faced for the successful application of XL-MIMO is the fact that, as the number of elements of the arrays --and consequently, the dimension of the XL-MIMO matrix-- increases, so does the complexity of the associated baseband processing for tasks such as MIMO precoding and detection. There is thus a need to design efficient signal processing tools that deal with large-dimensional problems and are scalable \cite{amiri2019message}. In this context, machine learning (ML) and, in particular, deep learning methods emerge as an almost indispensable tool to revisit complex scenarios.  Deep neural networks (DNNs) and especially convolutional neural networks (CNNs) have achieved considerable success in countless applications \cite{Bishop,Goodfellow2016}. Recently, they have attracted lots of attention in the field of wireless communications. Different problems in the physical layer such as rate maximization, modulation schemes, power control are investigated using machine learning techniques \cite{Deep_physical}.

In this work, we explore the application of DNNs to the problem of spatial user mapping  and MIMO precoding in the downlink of XL-MIMO systems. Motivated by the fact that different parts of the array may experience different propagation environments due to the large physical dimensions of XL-MIMO arrays, we pose a model for the XL-MIMO channel in which the signal received from (or transmitted to) a given user concentrates most of its energy over a limited portion of the array, which we call the visibility region (VR) of the user \cite{anum_nons}. Under such conditions, we realize that most of the gain from the XL-MIMO system can be achieved by precoding the signals transmitted to a given user over just a subset of the array elements belonging to its VR, rather than the whole array. Our goal is then to find the optimal subset of antennas for each user, given a fixed zero-forcing (ZF) precoder design and under a constraint on the maximum number of antennas that can be used for each of the users. Since solving the problem optimally requires combinatorial complexity, we instead train a deep CNN structure to perform the task. Once the network is trained, it provides a  competitive solution for the user mapping problem in comparison to the current state-of-the-art, as shown in our simulation results, the trained CNN is able to produce the optimal solution with high success rate and with negligible loss in terms of average sum-rate with respect to optimal ZF precoding over all the array elements.

\subsection{Related Works and Contribution}

%Applying DNNs in the research area of the MIMO systems has attracted attention due to complexity of many open problems like beamforming, AS and etc. 
Several works have already explored the application of DNN's to problems arising in MIMO communications. For example, the authors of \cite{deep_miso} use a deep learning framework to solve problems such as signal to noise and interference ratio (SINR) balancing, power minimization and sum-rate maximization in a MISO downlink system with $6$ antennas.  An application of ML for selecting antenna sets in MIMO systems is presented in \cite{ML_AS}, where the authors use well-known ML tools in a small MIMO system with $8$ antennas. The work in \cite{Deep_radar} focuses on antenna selection (AS) problem in a multi-antenna ($20$ antennas) system and constructs a CNN as a multi-class classification framework where each class designates a different subarray. 
% A DL based channel prediction is proposed in \cite{Deep_FDD} where they try to find a correspondence between uplink and downlink channel in frequency division duplex (FDD) massive MIMO with $32$ elements.

In all the aforementioned works, however, the system dimensions are limited and the effects of extra-large arrays are not explored. Contrary to this, we propose in this article a CNN based solution for the SUM problem in XL-MIMO arrays, where the signal received from (or transmitted to) a given user has most of its energy confined in the user's VR. Classical solutions to the AS problem such as \cite{MIMO_AS} are not practical for XL-MIMO, as the complexity explodes due to the increased channel dimensions. In our design, we keep the computational complexity contained by exploiting the users' VRs to distribute the SUM problem into a set of identical CNN's that can be run in parallel for each of the users. The simulation results show that our proposed solution performs very closely to optimal ZF-precoding over the whole antenna array, avoiding the exhaustive search technique. To the best of our knowledge, this is the first work applying deep learning methods to XL-MIMO systems.

%\cite{xlmimo_GC} tries to utilize the information about VRs of the users to carry out a simple distributed receiver. However, a central processing unit (CPU) is needed to convey information between these distributed units.
%Classical solutions to AS problem use heuristics to search over all possible antenna combinations maximizing an objective function \cite{MIMO_AS}. Such solutions does not guarantee optimality and due to their small searching space size, are not practical for our system model with hundreds of antennas.
 
%The problem that we are trying to solve has not been addressed before in the context of the XL-MIMO systems. To the best of our knowledge it is the first work applying ML to XL-MIMO and it uses distributed units working in parallel to exploit the features of VRs and make it practical for implementation.

% $*$paper structure
\section{System Model}

\begin{figure}
	\centering
 	\includegraphics[width=1\linewidth]{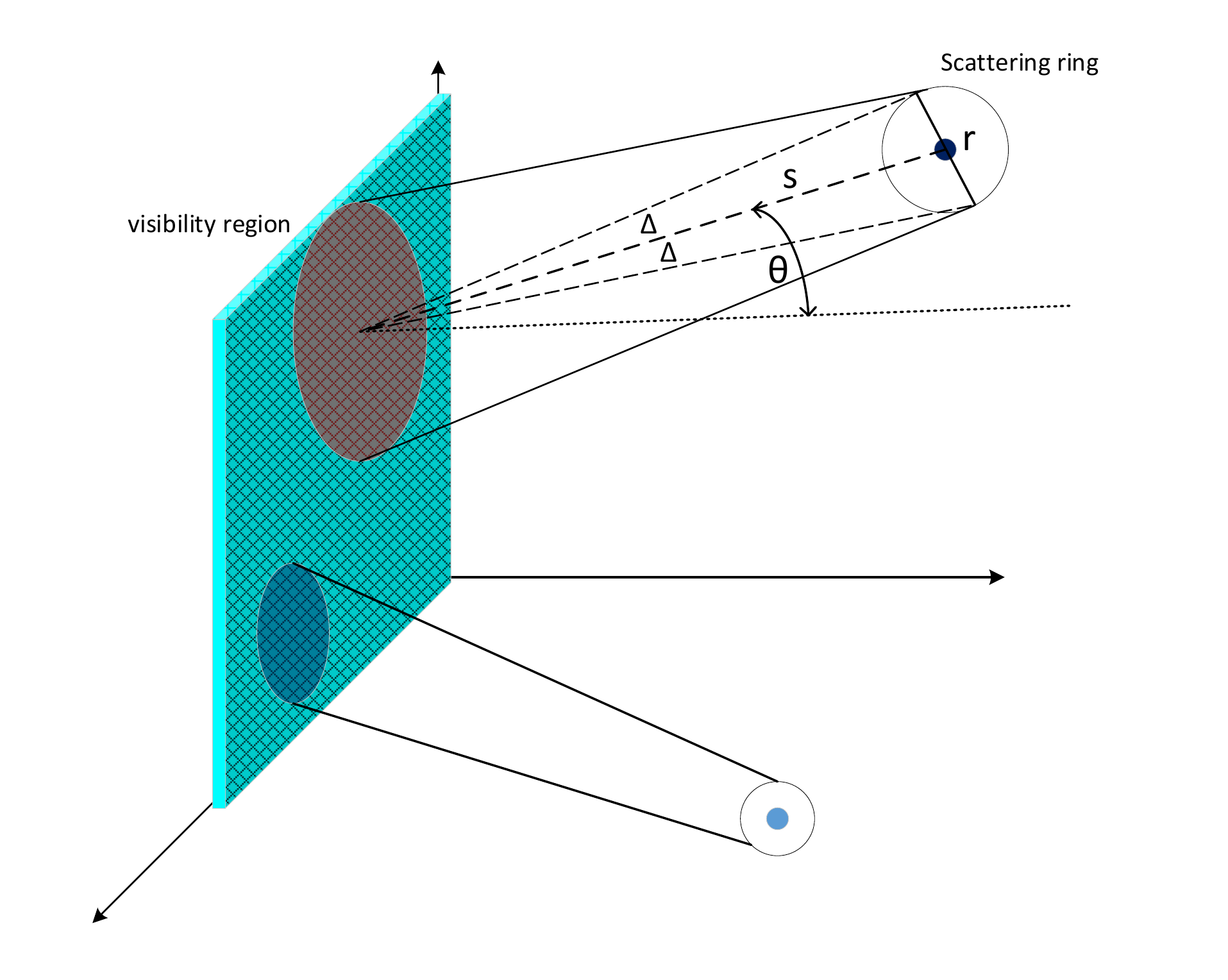}
	\caption{ \small An example of XL-MIMO array with spatial non-stationary regions along the array. Each user has a specific visibility region according to the channel conditions.
	}
	\label{fig:ex1}
	\vspace{-0.5cm}
\end{figure}

Let $M$ and $K$ denote the number of antennas and simultaneously active users, respectively.
We assume narrow-band transmissions; 
$\mathbf{x}\in\mathbb{C}^K$ denotes the vector of complex input symbols, $\mathbf{H}\in\mathbb{C}^{M\times K}$ is the complex channel matrix, $\mathbf{F}\in\mathbb{C}^{M\times K}$ is the MIMO precoding matrix and ${n}_k\sim \mathcal{CN}(0,\sigma_n^2)$ is the AWGN. We model the received baseband signal for user $k$ with ${y}_k\in\mathbb{C}$ as follows:
\begin{align}\label{eq:general_model}
{y}_k=\mathbf{h}_k^H\mathbf{F}\mathbf{x}+n_k.
\end{align}
where $(.)^H$ shows conjugate transposition and $\mathbf{h}_k$ denote the $k$-th column of $\mathbf{H}$, corresponding to user $k$.
%Also, channel matrix can be written as
%$ \mathbf{H}=[\mathbf{h}_1\:\mathbf{h}_2\dots\mathbf{h}_K] $ with $ \mathbf{h}_k $ denoting the channel for user $ k $.
In this work, we adopt the following channel  model \cite{larsson_energy}:
\begin{align}\label{eq:ch_model}
   \mathbf{h}_k=\sqrt{w_k} \bar{\mathbf{h}}_k,
\end{align}
where $w_k$ captures the effect of large scale fading which in turn is a function of the distance of the user from the array, denoted with $s_k$, and the propagation properties of the environment; here, we employ the following simplified propagation model \cite{Tse:2005:FWC:1111206}:\vspace{-0.3cm}
\begin{align}
w_k=\beta_k s_k^{\alpha},
\end{align}
where $\beta_k$ is a attenuation coefficient \cite{Tse:2005:FWC:1111206} and $\alpha$ is the pathloss exponent.
 $\bar{\mathbf{h}}_k \sim \mathcal{CN}(0,\mathbf{R}_k) $ accounts for fast fading in a non- line of sight scenario, with $\mathbf{R}_k$ being a symmetric positive semi-definite channel covariance matrix. We can represent channel vectors $\bar{\mathbf{h}}_k$ using Karhunen-Loeve expansion as \vspace{-0.2cm}
  \begin{align}\label{eq:KL_representation}
     \bar{\mathbf{h}}_k=\mathbf{U}_k \mathbf{\Lambda}_k^{\frac{1}{2}}\mathbf{z}_k ,
 \end{align}
 where $\mathbf{z}_k\in\mathbb{C}^{\zeta\times 1}\sim \mathcal{CN}(0,\mathbf{I})$, $\mathbf{\Lambda}_k$ is an $\zeta\times \zeta$ diagonal matrix with dominant eigenvalues and $\mathbf{U}_k\in\mathbb{C}^{M\times \zeta} $ is the tall unitary matrix of the eigenvectors of $\mathbf{R}_k$
corresponding to the $\zeta$ dominant eigenvalues.

Using the well-known \textit{one-ring} model \cite{one_ring} to define $\mathbf{R}_k$,  the correlation between the channel coefficients of antennas $p$ and $q$ is given by\vspace{-0.3cm}
\begin{align}\label{eq:R_def}
    [\mathbf{R}_k]_{p,q}=\frac{1}{2\Delta}\int_{-\Delta}^{\Delta}\exp{\bigl(j\mathbf{f}(\alpha+\theta)(\mathbf{u}_p-\mathbf{u}_q)\bigr)}\text{d}\alpha,\vspace{-0.2cm}
\end{align}
 where $\mathbf{f}(\nu)=-\frac{2\pi}{\lambda}\left(\cos(\nu),\sin(\nu)\right)$ is the wave vector with angle of arrival of $\beta$, carrier wavelength of $\lambda$ and $\mathbf{u}_p,\mathbf{u}_q \in \mathbb{R}^2$ are the position vectors of the antennas $p,q$ within the VR of user $k$.  $\Delta$ is angular spread which is $\Delta\approx \arctan(\frac{r}{s})$, with $r$ standing for the ring of scatterers radius. Also, $\theta$ is the azimuth angle of user $k$ with respect to antenna array (See Fig.~\ref{fig:ex1}). When either of the antenna indices $p,q$ is outside the VR for user $k$, we have $[\mathbf{R}_k]_{p,q}=0$. We use a uniform linear array (ULA) configuration and assume that the scattering rings are distributed uniformly in front of the array.
 
 In this paper we consider the downlink (DL) data  transmission  and the target of SUM is to  find the best antenna set for each of the users to send the signals. Due to the reciprocity of DL and uplink (UL) channels in time division duplex (TDD), each VR represents the region of the array that contains substantial portion of the user's energy, e.g. $95\% $ of total received energy, in UL transmission. Therefore,  
  one obvious consequence of having VR for each of the users is that the optimal SUM of each user should be inside its VR because the signal outside it is very weak and negligible. As it can be seen in Fig.~\ref{fig:ex1}, considering the antennas within VR of one user in the one-ring model results in dividing the array into binary regions. In order to model the distribution of the VRs we use a uniform distribution over the array for the VR centers and a parametric uniform distribution for the length of VRs, i.e. $\text{VR}_{length}\sim U(l_1,l_2)$.
  
  In order to formulate the precoder and the signal to noise and interference (SINR) of transmitting DL signal only on a subset of antennas or \textit{processing window} for each user, we define $\mathbf{H}_T$ as truncated channel. $\mathbf{H}_T$ only contains the elements of $\mathbf{H}$ on the processing window for each of the users. So,
%   \begin{align}
      $\mathbf{H}_T=\mathbf{A}\odot\mathbf{H}$ where $(\odot)$ stands for the element-wise matrix product
%   \end{align}
  and $\mathbf{A}\in \{0,1\}^{M\times K}$ is a binary matrix with its $(m,k)$th element taking the value $1$ if the $m$th antenna is within the processing window of the user $k$, and $0$ otherwise. 
  Thus, the truncated ZF beamformer is
  \begin{align}\label{eq:truncated_F}
      \mathbf{F}=\sqrt{\frac{P_T}{\text{trace}(\mathbf{P}(\mathbf{H}_T^H\mathbf{H}_T)^{-1})}}\mathbf{H}_T(\mathbf{H}_T^H\mathbf{H}_T)^{-1}
  \end{align}
% The resulting mean square error is
% \begin{align}
%     \text{MSE}=(\mathbf{F}\mathbf{H}-\mathbf{I})\mathbf{P}_x(\mathbf{F}\mathbf{H}-\mathbf{I})^H+ \sigma_n^2 \mathbf{F}\mathbf{F}^H
% \end{align}
where $\mathbf{P}=\text{diag}(P_{x_1},P_{x_2},\cdots,P_{x_K})$ is a diagonal matrix with user signal powers, i.e. $P_{x_k}=\mathbb{E}\{x_kx_k^H\}$ and $\mathbb{E}$ stands for expectation operator; $P_T$ is the total transmit power. Finally the SINR for user $k$ is equal to
\begin{align}\label{eq:SINR}
\gamma_k=\frac{|\mathbf{h}_k^H\mathbf{f}_k|^2P_{x_k}}{\sum_{k\prime\neq k} |\mathbf{h}_k^H\mathbf{f}_{k\prime}|^2P_{x_{k\prime}} + \sigma_{n_k}^2}.
\end{align}
%\begin{align}\label{eq:SINR}
%    \gamma_k=\frac{\rho_k}{\text{trace}(\mathbf{P}(\mathbf{H}_T^H\mathbf{H}_T)^{-1})}\qquad \text{and} \qquad \rho_k=\frac{P_{x_k}}{\sigma_{n_k}^2}.
%\end{align}
%The second term is pre-processing SNR of each user.

\section{Spatial User Mapping Problem Formulation}

While the truncated ZF precoder in \eqref{eq:truncated_F} does not guarantee interference free reception, as signals sent from antennas outside the mapping window of user $k$ may still interfere user $k$'s reception, it has the advantage that the precoder for each user can be calculated based on a much smaller channel matrix: specifically, for computing user $k$'s precoder, one needs only to consider those columns of $\mathbf{H}_T$ which contain non-zero elements within user $k$'s mapping window. Thus we expect that, if the mapping window of the users is chosen near-optimally, we obtain a noticeable complexity reduction at the expense of only a slight performance degradation with respect to the non-truncated ZF precoder.

To this end, we formulate the SUM problem as the selection of $N_{max}$ antennas for each of the users that maximizes the system's sum-rate with the precoder \eqref{eq:truncated_F}, i.e.:
\begin{subequations}
\begin{align}
    &\max_{\{\mathbf{a}_k\}} \sum_{k=1}^K \log(1+\gamma_k)\\
    &\text{s.t.}\quad ||\mathbf{a}_k||_1=N_{max},\quad {k = 1,2,\dots,K}
\end{align}
\label{eq:primary_obj}
\end{subequations}
where $\gamma_k$ is the SINR for user $k$, $\mathbf{a}_k$ is $k$th column of $\mathbf{A}$, $||\mathbf{a}||_1$ is the norm $1$ of $\mathbf{a}$. %Note that in this work we fix the beam-forming method to ZF and other beamformers are left for future work.

In general, solving \eqref{eq:primary_obj} needs checking all the combinations of the antennas for all the users to see which one maximizes the sum-rate. The total number of the combinations is $K\frac{M!}{N_{max}!(M-N_{max})!}$ which is very large in our system. Therefore sub-optimal solutions were presented to use heuristic algorithms to locate best antenna sets to be selected \cite{MIMO_AS}. On the other hand, the computational cost of these methods is huge and is  impractical as $M$ and $K$ grow very large. Instead of solving the optimization directly, we propose in the next section a DNN structure which, once successfully trained,can produce acceptable solutions while providing a useful way of rethinking the precoder design. 

Having  a  XL-MIMO  array  provides  a  great  spatial  resolution to distinguish users’ signals better. For instance, users interference  is  limited  to  their  VR  overlaps  and  users  with no  VR  intersection  enjoy  interference  free  environment.  Since  ZF  solves a set of dependent linear equations  by backward substitution algorithm, it treats independent equations as a new linear system and solves them separately.  Thus,  extending it  to  the  concept  of  VR  in  XL-MIMO,  recovering  the  signal within  one  VR  is  possible  over  the  superposition  of  VRs  of the interfering users (see Fig. 2). We call this subset of users and antennas \textit{effective window}, and discuss its importance on the implementation of distributed CNNs in the next section.

\begin{figure}
	\centering
 	\includegraphics[scale=0.55,trim={0.6cm 0.6cm 0.6cm 0.6cm },clip]{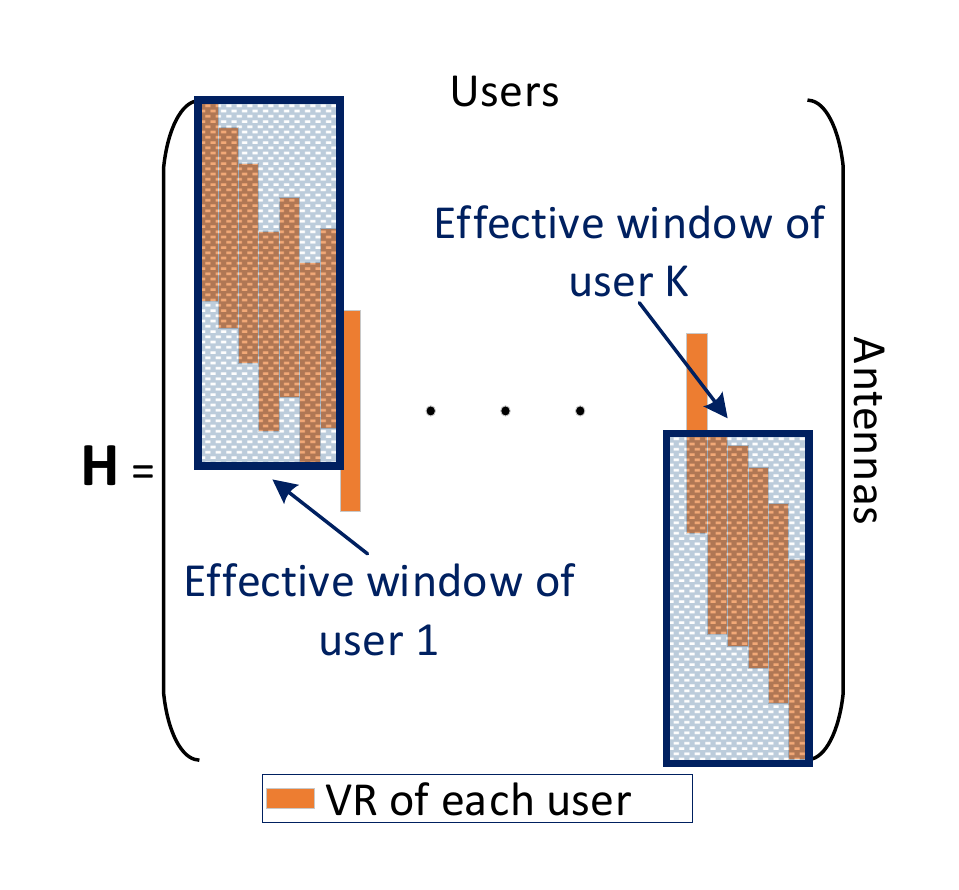}
	\caption{ \small An illustrative example of effective window concept. Only the interfering users affect one users signal and it is sufficient to consider only this part to extract the necessary information.
	}
	\label{fig:ex2}
\end{figure}

\section{CNN Deep Learning }

\vspace{-0.1cm}

As mentioned in the previous sections, the large dimensions of the problem prevents us from using simple learning methods such as K-nearest neighbors algorithm \cite{ML_AS}, which are impractical options for SUM in our context. As an alternative, convolutional neural networks have shown great performance in dealing with the problems in physical layer \cite{Deep_physical}. Especially when size of the optimization parameters  is very large, there exist an appropriate CNN with enough depth and training data that can extract all the features perfectly. In this section we present our CNN's layer architecture and their goal.

\begin{figure*}
	\centering
 	\includegraphics[scale=0.4]{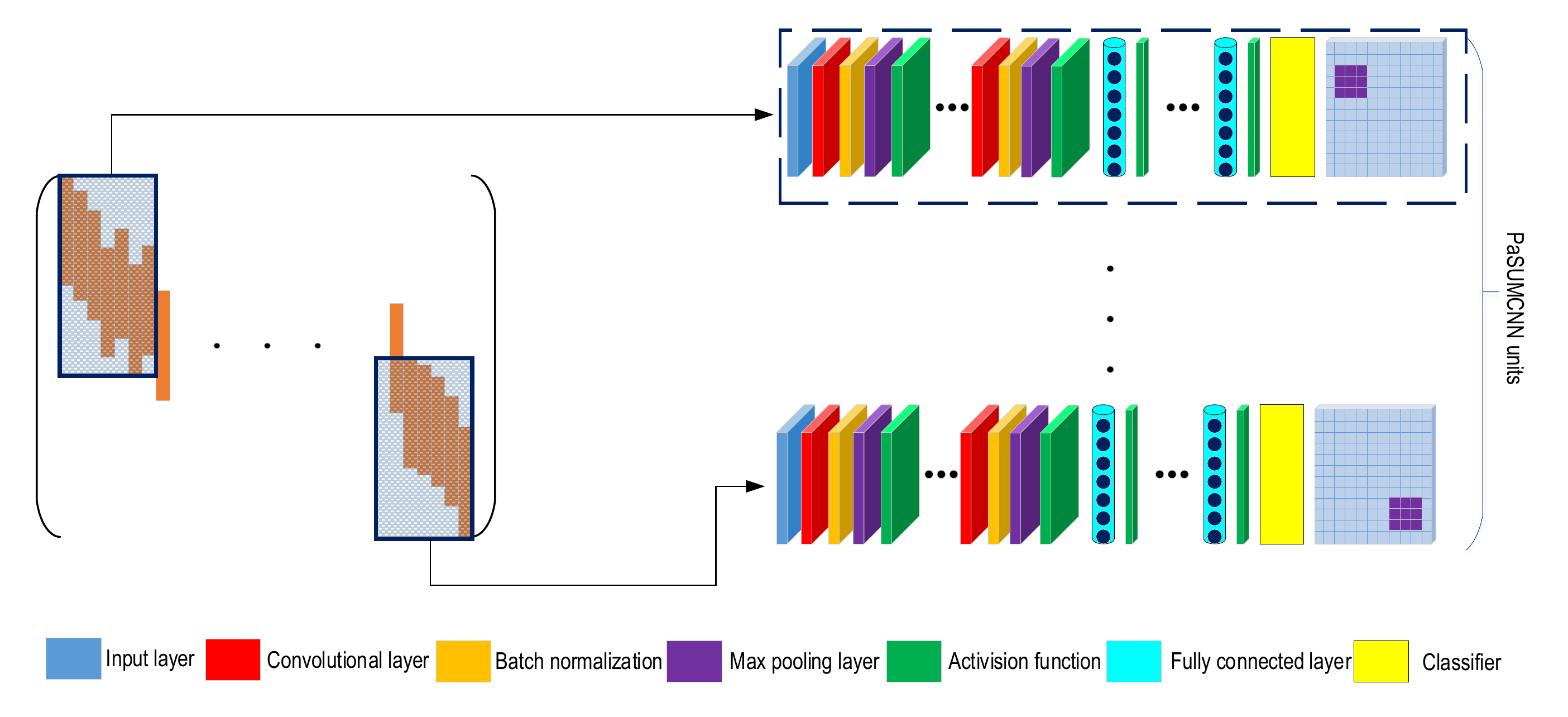}\vspace{-0.3cm}
	\caption{ \small Schematic of  PaSUMCNN  in distributed XL-MIMO. Each of these  PaSUMCNN  units are fed by the effective window of the corresponding user.
	}
	\label{fig:cnn}\vspace{-0.3cm}
\end{figure*}

\vspace{-0.1cm}

\subsection{Network Architecture}
The dashed box in Fig.~\ref{fig:cnn} shows the basic CNN design for the SUM problem. It consists of several layers starting with the input layer. To the best of our knowledge, current deep learning tools only work with real-valued data and therefore we enhance our original channel matrix data $\mathbf{H}$ by adding other dimensions. Input data has a dimension of $M'\times K'\times 3$ where $M'$ and $K'$ are the antenna elements and the users present in the effective window, respectively. The third dimension contains the real, imaginary and phase of the entries.

\textit{Convolutional layers (CL)} create several convolution kernels to be convolved with their input. These layers try to extract different features at each of of their kernels and are vastly used for image processing applications. Similar to images, elements of the channel matrix $\mathbf{H}$ have many similarities with their neighbor elements and we use this principle to implement two-dimensional  CLs. Otherwise, one should put a \textit{fully connected (FC) layer} with enormous number of neurons to capture this information. The latter solution is more complex due to its size and needs more training data. 

Batch normalization (BN) layers are placed after each CL for normalization and scaling their input.
% they subtract the batch mean and then divide it by the batch standard deviation
 They have two learnable parameters:  \textit{offset} and \textit{factor} (See Table \ref{tab:tab1}). BNs 
help in avoiding over-fitting, accelerating convergence and reducing the sensitivity to the initialization of the weights \cite{deep_miso}. Then, \textit{Max pooling (MP) layers} are placed to reduce the dimension of BN's output by only keeping the maximum element within their specific pooling window.

Most importantly the nonlinear units known as activation functions (AF), which are the distinct difference between conventional linear systems, are used after MP and FC layers. Here, we use rectified
linear units (ReLU) where the ReLU$(x) = \max(x, 0)$.
FC layers are well-known hidden layers in the neural networks.  
Right before the output layer we put a \textit{Flatten layer} which is a FC layer with number of the neurons equal to the number of the classes. This layer prepares the output for the \textit{Soft-max layer} which calculates the probability distribution of the classes.

In order to optimize \eqref{eq:primary_obj} using DNNs, similar to \cite{Deep_radar} we selected classification problem where each class stands for a set of $N_{max}$ antennas indices with total number of $\mathbb{C}$ classes. Since the VR for each of the users contains contiguous antenna elements over a region of the array and also to keep the complexity very low we assume these $N_{max}$ antennas are adjacent. These $\mathbb{C}$ classes are places in a uniform way to cover all the array elements. Then, we use the negative log-likelihood or cross-entropy loss function for the classifier. The accuracy of the classification is another metric that we use for evaluating the network's performance. We define accuracy as\vspace{-0.2cm}
\begin{align}
    \text{Accuracy} (\%)=\frac{T_s}{T}\times 100
\end{align}
where $T$ is the total number of samples in the training/evaluation datasets and $T_s$ is the number of samples for which the classifier successfully selects the correct class. 

\subsection{Parallel CNNs}
The most important obstacle for extending MIMO array size is the computational complexity growth. One way to deal with this burden is to employ distributed processing units \cite{xlmimo_GC}. One drawback of these methods compared to the central processing is to find the best solution to distribute tasks between the units. Inspired by the fact that sufficient information for detecting each user lies in its effective window region, we proposed parallel spatial user mapping convolutional neural networks ( PaSUMCNN ) for our problem. As shown in Fig.~\ref{fig:cnn} each of these identical networks only processes the effective window of the desired user. This method has the following advantages:
\begin{itemize}
    \item Small sized CNNs can be used since the size of the feature domain is limited.
    \item These small sized CNNs need ordinary distributed processing units that are not costly compared to ultra fast computing systems.
    \item All the units work in parallel and there is no extra delay for information exchanges between the units.
    \end{itemize}

\subsection{Spatial User Mapping Method for CNN Training}
In this section we describe the training data design process and the main assumptions for the channel model and the training phase.
\\
Algorithm~\ref{alg} describes the training data generation procedure. After initializing with simulation and system parameters, it generates all the possible antenna selections for all the users  in step $1$. A set of $N_{max}$ antenna indices is allocated for each user and users can have similar sets as well. 
% In order to keep the size of searching space limited and also use some of the features of VR distributions we assume consecutive antenna indices. This comes from the fact that VRs in our channel model are continuous. We discuss the effect of this assumption in section \ref{sec:sim}.  

In the next step, a channel matrix is generated and is used for calculating SINR \eqref{eq:SINR} and the sum-rate \eqref{eq:primary_obj}. Afterwards, the best configuration maximizing the sum-rate is selected as the solution for the SUM problem. The last step, concatenates the channel matrices and the labels for all of the samples. There is also one hidden step before feeding these training data to the CNNs and there we normalize and scale the channel matrices (same as BN layer) to avoid effects of using biased data. The normalization and scaling has the following form:
\begin{align}\label{eq:norm}
    [\mathbf{H}]_{ij}^{\text{new}}=\frac{[\mathbf{H}]_{ij}^{\text{old}}-\frac{1}{M}\sum_i[\mathbf{H}]_{ij}^{\text{old}}}{\max_i [\mathbf{H}]_{ij}^{\text{old}}-\min_i [\mathbf{H}]_{ij}^{\text{old}}} .
\end{align}

\begin{algorithm}[t]
	\SetAlgoLined
	\KwResult{Labeled training and validation data for each of the  PaSUMCNN  units $\mathcal{T}_k\:,\:k=1,\cdots K$}
	\emph{Initialize:} 
	 $M$,  $K$, $N_{max}$, $\mathbb{C}$, number of samples $\mathbb{I}$, $\sigma^2_n$,  $\mathbf{P}$,  $\beta,\:\mathbf{d}\:\text{and }\alpha$.
	\\
	1. Calculate all possible antenna selection space $\mathbf{Q}$, where  $|\mathbf{Q}|=K^{\mathbb{C}}$.\\
\For{$i = 1$ to $\mathbb{I}$ }{

2. Generate channel matrix $\mathbf{H}$ using \eqref{eq:ch_model}.

\For{$q = 1$ to $|\mathbf{Q}|$ }{

3. Calculate $\mathbf{H}'=\mathbf{H}^q_T$ which is channel matrix over antenna selection possibility $q$.

\For{$k = 1$ to $K$ }{

4. Calculate $\gamma_k^{(q,i)}$ using \eqref{eq:SINR} with proper $\mathbf{F}$ given by \eqref{eq:truncated_F} and $\mathbf{H}'$.

		}
		
5. Calculate the sum-rate using \eqref{eq:primary_obj} and $\gamma_k^{(q,i)}$s.
		
		}
		
6. Labeling: find $\mathcal{D}_{k,i}=\text{argmax}_q \sum_{k=1}^K \log(1+\gamma_k^{(q,i)} )\:\forall{q}$.		
		}
		
7. Concatenation: concatenate training data and corresponding labels as $\mathcal{T}=\{\mathbf{H}^{(i)}|\mathcal{D}^{(i)}\}\:,\:\forall{i\in \mathbb{I}}$.

	\caption{\small Training data generation.}
	\label{alg}
\end{algorithm}

\section{Simulation results and discussion }
\label{sec:sim}
In this section, the simulation process is described. We also address the challenges during the development of  PaSUMCNN  and the provided solutions.

\begin{table}
\centering
\caption{CNN architecture parameters in detail.}
\label{tab:tab1}
% \begin{tabular}{|p{2.45cm}|c|p{2.5cm}|}
\begin{tabular}{|c|c|c|}
\hline
\bf{Layers} & \bf{Activations} & \bf{Learnables} \\\hline	
Input  & $128 \times 4\times3$   & -          \\ \hline
   {Conv. 2D }     &    \multirow{2}*{ $97 \times 3\times256$   }     &     weights : $3\times 2^{14} $      \\
  $256\,\text{\texttimes}\, [32 \times 2\times3]$&&bias : $ 256$ \\\hline
  
{Batch norm. }     &    \multirow{2}*{ $97 \times 3\times256$   }     &     offset : $ 256$      \\
  $256\,\text{channels}$&&scale : $ 256$ \\\hline
  
{Max pooling }     &   { $72 \times 3\times256$   }     &     -     \\\hline
  
{ReLU }     &   { $72 \times 3\times256$   }     &     -     \\\hline

{Conv. 2D }     &    \multirow{2}*{ $41 \times 2\times32$   }     &     weights : $2^{19} $      \\
  $32\,\text{\texttimes}\, [32\! \times\! 2\!\times\!256]$&&bias : $ 32$ \\\hline
  
{Batch norm. }     &    \multirow{2}*{ $41 \times 2\times32$   }     &     offset : $ 32$      \\
  $32\,\text{channels}$&&scale : $ 32$ \\\hline  
 
{Max pooling }     &   { $16 \times 2\times32$   }     &     -     \\\hline 

{ReLU }     &   { $16 \times 2\times32$   }     &     -     \\\hline  

{Fully connected }     &    \multirow{2}*{ $1024$   }     &     weights : $2^{20}$       \\
  $1024$&&bias : $ 2^{10}$ \\\hline

{ReLU }     &   { $1024$   }     &     -     \\\hline

{Fully connected }     &    \multirow{2}*{ $64$   }     &     weights : $2^{16}$       \\
  $64$&&bias : $ 64$ \\\hline

{ReLU }     &   { $64$   }     &     -     \\\hline

{Fully connected }     &    \multirow{2}*{ $16$   }     &     weights : $1024$       \\
  $16$&&bias : $ 16$ \\\hline
  
{ReLU }     &   { $16$   }     &     -     \\\hline
 
{Flatten }     &    \multirow{2}*{ $\mathbb{C}$   }     &     weights : $\mathbb{C}\times 16$       \\
  $\mathbb{C}$&&bias : $\mathbb{C}$ \\\hline 

{Softmax}     &   { $\mathbb{C}$   }     &     -     \\\hline

\end{tabular}

\end{table}

Table \ref{tab:tab1} summarizes the layer parameters of each CNN unit depicted in Fig.~\ref{fig:cnn}. We use mini-batches of size $250$ in the training process with a learning rate of $4\times10^{-4}$. 

Our designed network has two  Convolutional blocks including a set of CL-BN-MP-AF  in the input. Adding more blocks had no effect on improving the performance of the system with the same amount of training data. 
All the simulation parameter values are mentioned in Table \ref{tab:tab2}.

\subsection{Managing the over-fitting problem}
 Usually,  we divide our labeled data into two parts: a major part for training the network and a minor part to validate the accuracy of the trained network.
 We used $20\%$ of the labeled samples for the validation part.
 One of the common issues that happens during the process of training a machine is the \textit{over-fitting} problem, when the networks adapts itself more to the variability of the training data and it fails to act properly with a new data set. We use three methods to deal with this issue \cite{Bishop}:  

\begin{itemize}
    
    \item \textbf{Normalizing data}: We use BN layers and also \eqref{eq:norm} to avoid any bias in the input data.
    \item \textbf{Dropout layer}: These layers randomly discard some of the neurons of FC layers according to a probability $p$ at each epoch of the training. This helps in dropping the neurons with dominant weights in the network  allowing the prediction of new data sets.
    \item \textbf{Regularization factor}: Adding a penalty term to the loss function of the network assists in bypassing the over-fitting. We use a first order penalty term in our network with regularization factor $\epsilon$.
 
 \end{itemize}
 
\begin{figure}
	\centering
 	\includegraphics[width=1\linewidth,trim={2.2cm 0 2.2cm 0 },clip]{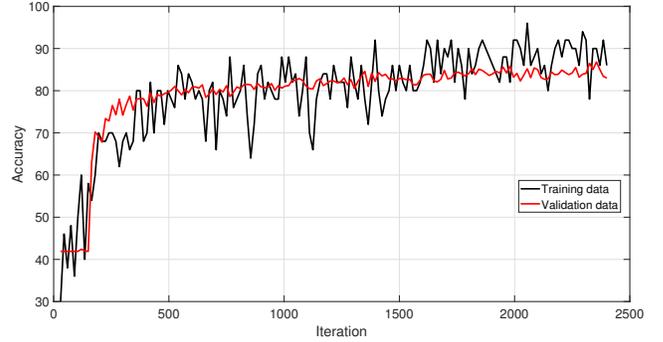}
	\caption{ \small Convergence of the training phase of the  PaSUMCNN  method.
	}
	\label{fig:sim3}
\end{figure}

Using these methods, the convergence behaviour of accuracy in the training phase of the network is presented in Fig.~\ref{fig:sim3} for both of the training and validation data. Both curves  reach $80\%$ accuracy levels at early stages of the training ($1/5$th of the training period). Thus, we can compromise a bit on the accuracy of the model to have a faster training phase for delay sensitive applications. 

\begin{table}
\centering
\caption{Simulations parameters in detail.}
\label{tab:tab2}
% \begin{tabular}{|p{2.45cm}|c|p{2.5cm}|}
\begin{tabular}{|c|c||c|c|}
\hline
\bf{Variable} & \bf{Value} & \bf{Variable}& \bf{Value} \\\hline	

$M$    &    $256$  &   $K$   &  $9 $  \\\hline

$\mathbb{C}$    &    $12$  &   $N_{max}$   &  $40 $  \\\hline

$\mathbf{P}$    &    $\mathbf{I}$  &   $\mathbb{I}$   &  $5000 $  \\\hline

$\beta$    &    $2$  &   $\alpha$   &  $3 $  \\\hline

$\lambda$    &    $2.6$GHz  &   Antenna spacing   &  $\lambda/2 $  \\\hline

$(l_1,l_2)$ & $(50,100)$  &  $\zeta$   &  $M/4 $  \\\hline

$p$    &    $0.6$  &   $\epsilon$   &  $0.0015 $  \\\hline

\end{tabular}

\end{table}

\subsection{Numerical Results}
\begin{figure}
	\centering
 	\includegraphics[width=0.95\linewidth,trim={2.2cm 0 2.2cm 0 },clip]{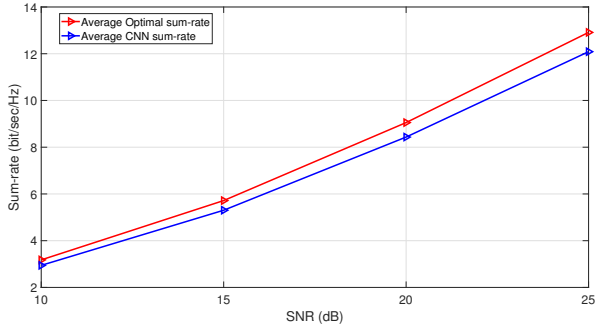}
	\caption{ \small Sum-rate of the optimal  and learning-based solutions.
	}
	\label{fig:sim1}
\end{figure}
Once the network is trained, we use it to perform SUM on newly generated channel matrices. The results are shown in Fig.~\ref{fig:sim1}, which compares the sum-rate of the optimal SUM (exhaustive search) and  PaSUMCNN  versus different pre-processing SNR values. As it can be seen the performance gap is very small and ranges between $6.4\%$ to $7.4\%$.
Moreover, our assumption on having consecutive antenna sets for the SUM solution space is not creating a huge difference here and we can claim that it is a valid simplifying assumption.  
This proves the near-optimal output of the  PaSUMCNN in the XL-MIMO systems. 
% It is worth mentioning that the computational complexity of the PASCNN is much smaller than the exhaustive search as demonstrated above. Having a much lower complexity will allow us to use this method in action. Training of the network occurs offline and at each large-scale fading coherence interval, then at each time we only need to feed the CSI and do a series of matrix multiplications within each of the networks.

\vspace{-5pt}

\subsection{Implementation aspects}
In this subsection we roughly compare the computational complexity of the  PaSUMCNN  and the exhaustive search method. For the  PaSUMCNN  method as demonstrated in Table \ref{tab:tab1}, the dominant number of weights is $2^{20}$ which controls the complexity of the multiplications for this method. It also has $K$ parallel units that each have the same complexity which roughly gives $K\times 2^{20}$ computations per SUM and it is independent of the number of the classes $\mathbb{C}$.  On the other hand, in exhaustive search we look into $K^\mathbb{C}$ possibilities and for each of them the matrix inversion of ZF needs $K^3$ operation giving roughly $K^{\mathbb{C}+3}$ computations per AS. As an example and with our simulation parameters, the complexity gain of  PaSUMCNN  is approximately $\frac{K^{\mathbb{C}+3}}{K\times 2^{20}}=2\times10^7$.
It is worth mentioning that this complexity gain of the  PaSUMCNN  creates a great opportunity to implement this method for practical applications. Training of the network occurs offline, then at each time we only need to feed the CSI and execute a series of matrix multiplications within each of the networks.

% \begin{figure}
% 	\centering
%  	\includegraphics[width=1\linewidth]{acc_bar_23apr.eps}
% % 	\includegraphics[width=1\linewidth]{Drawings/VLA_nonS3.eps}
% 	\caption{ \small Comparison of the accuracy level of training versus validation data.
% 	}
% 	\label{fig:sim2}
% \end{figure}

\section{Conclusions}

\vspace{-0.1cm}

In this work, we have presented a deep learning based solution to optimize the operation of XL-MIMO arrays. In particular, our solution seeks an optimal mapping of users to parts of the array, such that the DL sum-rate using a truncated ZF precoder is maximized. Our results show that a properly trained network is able to solve this problem nearly as well as an optimal mapping algorithm, while proposing a competitive solution with respect to the current state-of-the-art solutions that only consider small sized systems.

The results presented in this work lead us to two general conclusions: 1) increasing the size of current massive MIMO arrays is one of the most promising avenues to continue boosting the spectral efficiency of current and future wireless systems, and 2) the recent developments in automated learning algorithms, such as deep learning, enable a new research direction in the communication systems to model and solve complex problems. We conjecture that the combination of these two aspects will be one of the leading trends in the developments of wireless communications in the future years.

%This work considers a XL-MIMO system, a massive MIMO system with an extremely large physical size that goes beyond the dimensions of traditional MIMO systems. Because of this large dimensionality, the optimization of an XL-MIMO system leads to solutions with prohibitive complexity when relying on conventional optimization tools. Generally, we advocate the use of machine learning to face the challenges posed by the design of a XL-MIMO system, ranging from physical layer optimization problems such as beamforming or resource allocation.  
%As a first effort towards establishing the benefits of machine learning, we propose a cross-layer design for the downlink of a multi-user setting with linear pre-processing, where the goal is to select a limited processing area per user, i.e. a small portion of the array that contains the beamforming energy to the user. Our solution relies on deep learning with a distributed architecture that is built to manage the large system dimension. This architecture contains one network per user working in parallel and exploits specific non-stationary properties of the channel along the arrays.  
%We show performance results that are very close to the optimal one while providing low computational complexity. 

%\nocite{*}

\bibliographystyle{IEEEtran}

\end{document}